\documentclass[aps,prl,reprint,groupedaddress]{revtex4-1}

\usepackage{bm}
\usepackage{amsmath}
\usepackage{amssymb}
\usepackage{graphicx}
\usepackage{color}

\begin{document}


\title{Microscopy for Atomic and Magnetic Structures Based on Thermal Neutron Fourier-transform Ghost Imaging}


\author{Kun Chen}
\email[]{kunchen@siom.ac.cn}
\affiliation{Key Laboratory for Quantum Optics and Center for Cold Atom Physics, Shanghai Institute of Optics and Fine Mechanics, Chinese Academy of Sciences, Shanghai 201800, China}
\author{Shensheng Han}
\affiliation{Key Laboratory for Quantum Optics and Center for Cold Atom Physics, Shanghai Institute of Optics and Fine Mechanics, Chinese Academy of Sciences, Shanghai 201800, China}


\date{\today}

\begin{abstract}
We present a lensless, Fourier-transform ghost imaging scheme by exploring the
fourth-order correlation function of spatially incoherent thermal neutron waves.
This technique is established on the Fermi-Dirac statistics and the
anti-bunching effect of fermionic fields, and the analysis must be fully quantum
mechanical.  The spinor representation of neutron waves and the derivation
purely from the Schr\"{o}dinger equation makes our work the first, rigorous,
robust and truly fermionic ghost imaging scheme.  The investigation demonstrates
that the coincidence of the intensity fluctuations between the reference arm and
the sample arm is directly related to the lateral Fourier-transform of the
longitudinal projection of the sample's atomic and magnetic spatial
distribution.  By avoiding lens systems in neutron optics, our method can
potentially achieve de Broglie wavelength level resolution, incomparable by
current neutron imaging techniques.  Its novel capability to image crystallined
and noncrystallined samples, especially the micro magnetic structures, will
bring important applications to various scientific frontiers.
\end{abstract}

\pacs{42.50.Ar, 61.05.Tv, 42.30.Va, 42.50.St}


\maketitle


Ever since the discovery of the Hanbury Brown and Twiss (HBT)
effect\cite{HBT-Nature1956a,HBT-Nature1956b}, the quantum statistical properties
of bosonic and fermionic fields have been intensively investigated.  It is found
that the intensity correlation in HBT measurements actually originates from the
photon bunching of thermal light sources.  In parallel, a distinctive
antibunching effect, with its deep roots in the Pauli exclusive principle,  was
predicted for
fermions\cite{Silverman-NuoCim1987a,Silverman-NuoCim1987b,Tyc-PRA1998}, and
subsequently observed in free thermal
neutrons\cite{Iannuzzi-PRL2006,Iannuzzi-PRA2011}, free
electrons\cite{Klesel-nature2002}, two dimensional electron gas of semiconductor
systems\cite{Henny-Science1999,Oliver-Science1999}, and free fermionic atoms
released from an optical lattice.\cite{Rom-nature2006}.

Over the past two decades the bosonic bunching property has led to an imaging
methodology, called quantum imaging or synonymously ghost imaging (GI), by
exploring the intensity correlation between split beams.  It was first realized
with entangled photon pairs generated by spontaneous parametric down conversion
from a nonlinear crystal\cite{Shih-PRL2001}.  Further developments demonstrated
that quantum entanglement was unnecessary\cite{Boyd-PRL2002} and thermal light
could also be employed\cite{HSS-PRL2004,Gatti-PRL2004}.  Moreover, such a
concept is directly expandable to the de Broglie waves of massive
particles\cite{Khakimov-Nature2016}.

Essentially a nonlocal imaging scheme, GI provides vast flexibility in optical
system design without requiring a lens system for focusing and magnification,
and has wide applications from long range
remote-sensing\cite{HSS-APL2012,Hardy-PRA2013} to short range
microscopy\cite{Aspden-Optica2015}.  Different from conventional methods, the
light field transmitted through or reflected by a sample is recorded only with a
non-spatially resolving detector (i.e., a bucket or point detector), while the
spatial profile is recorded by a high-resolution reference detector.  From the
Fourier-transform diffraction pattern collected at the Fresnel region, the
sample structure can be successfully reconstructed\cite{HSS-PRA2007}.  In latest
developments, hard x-ray GI was achieved experimentally by employing synchrotron
radiation\cite{HSS-PRL2016,Pelliccia-PRL2016}.  Theoretically the spatial
resolution of lensless Fourier-transform GI is only limited by the wavelength
and provides the potential to achieve atomic-resolution imaging of
noncrystalline samples using laboratory x-ray sources\cite{HSS-PRL2016}.  The
quantum waves of massive bosons were brought into GI as well.  Correlated
metastable helium atom pairs, extracted from a Bose-Einstein condensate, were
shown to be capable of generating a 2D imaging of a target with submillimeter
resolution\cite{Khakimov-Nature2016}.

As a major achievement of modern physics, thermal neutron scattering has greatly
enriched our understanding of atomic-scale structures and dynamical
properties of materials\cite{Squires1978,Lovesey1984}.  
With de Broglie wavelength at the same order of
interatomic distances in solids and liquids, carrying no net charge, and
participating nuclear scattering and magnetic scattering, thermal neutrons make
an ideal probe for detecting matter's atomic and magnetic
structures.  Among all subfields of neutron science, neutron attenuation
imaging, phase imaging and holography have been under intense
developement and created a variety of important tools\cite{Anderson2009}.  The
latest, state-of-the-art neutron microscopy, by employing
multi-layered Wolter mirrors, not only 
focuses the neutron beam onto the target, but also magnifies the
image\cite{NeutronWolterImaging2013}.  Tweenty micron resolution has been
achieved while a potential resolution of 1 micron is reachable if phase contrast
information is further incoporated, though still orders of magnitude larger than
the wavelength.

Thermal neutrons and hard x-rays have similar wavelength range.  As x-rays
are mostly scattered by the electric field of electrons, they are suitable for
studying atomic structures comprised of heavy elements, but incapable
of probing magnetic structures.  Complementary to x-rays, thermal neutrons are
sensitive to the nuclear scattering of light-elements and the magnetic
scattering of unpaired electrons.  Thus, a thermal neutron GI technique can
greatly improve the resolving power of seeing through matter's atomic and
magnetic structures.  However, due to the complexity in calculating the
fourth-order correlation function of neutron scattering processes, particularly
when involving neutron's magnetic moment, GI employing thermal neutron probe has
yet to be developed.  Here for the first time in literature, we present a
Fourier-transform imaging scheme for atomic and magnetic structures based on the
neutron quantum coincidence in a GI setup.

\begin{figure}
\includegraphics[width=\linewidth]{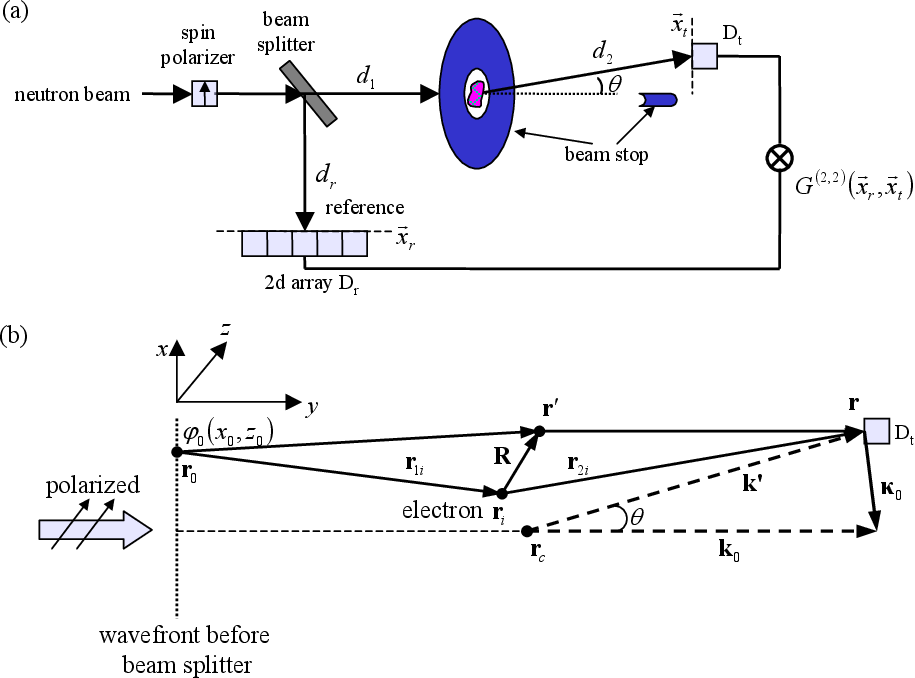}
\caption{Illustration of thermal neutron ghost imaging.  (a) Schematic
experimental setup.  A polarized (in $\hat z$ dirction), spatially incoherent 
thermal neutron beam is divided into a sample arm and
a reference arm.  An extra beam-stop shield is inserted around the sample to
remove the incident waves, while the sample is mounted in the opening window of the
screen. (b) The definitions of the coordinate system and the vectors
used in the calculations, the angles are exaggerated: $\mathbf{r}_0$ one point
on the source wavefront, $\mathbf{r}_i$ the $i$-th unpaired electron,
$\mathbf{r}$ the detector, $\mathbf{r'}$ a magnetic field point, $\mathbf{r}_c$
the system center.\label{fig1}}
\end{figure}

In GI, the particle flux is divided into two by a beamsplitter, one
reference arm and one sample arm (Fig.~\ref{fig1}a).  The coincidence rate at
these two detectors is proportional to the fourth-order correlation function of
the quantum fields,
\begin{equation}
G^{(2,2)}(\bm{\xi}_r,\bm{\xi}_t)=\left<\psi^\dagger(\bm{\xi}_r)\psi^\dagger(
\bm{\xi}_t)\psi(\bm{\xi}_t)\psi(\bm{\xi}_r)\right>,
\label{eq:g22}
\end{equation}
where $\left<\cdot\cdot\cdot\right>$ means the ensemble average, and we use
Greek symbol $\bm{\xi}=(x,z)$ to label the 2d transverse coordinate of the
wavefront.  $\bm{\xi}_r$ and $\bm{\xi}_t$ are for the reference and target
detectors, respectively.  The propagation of quantum field is governed by
\begin{equation}
\psi(\bm{\xi})=\int h(\bm{\xi},\bm{\eta})\varphi(\bm{\eta})d\bm{\eta},
\label{eq:psivarphi}
\end{equation}
where $\varphi$ and $\psi$ are the source and output fields, respectively;
$h(\bm{\xi},\bm{\eta})$ is the impulse response function, and $\bm{\eta}$ labels
the transverse coordinate of the wavefront at the source.  The statistical
properties of $\psi$ and $\varphi$ are bridged by the function
$h(\bm{\xi},\bm{\eta})$.  Theory has shown that on the source
wavefront\cite{Tyc-PRA1998,WangKG-PRA2009,Goodman1985}
\begin{eqnarray}
&G^{(2,2)}(\bm{\eta}_1,\bm{\eta}_1',\bm{\eta}_2,\bm{\eta}_2')
&=G^{(1,1)}(\bm{\eta}_1,\bm{\eta}_2')
G^{(1,1)}(\bm{\eta}_1',\bm{\eta}_2)\nonumber\\
&&\pm G^{(1,1)}(\bm{\eta}_1,\bm{\eta}_2)G^{(1,1)}(\bm{\eta}_1',\bm{\eta}_2'),
\label{eq:g22g11g11}
\end{eqnarray}
where $G^{(1,1)}(\bm{\eta},\bm{\eta'})$ is the second-order correlation function
of the source field, defined as
$G^{(1,1)}(\bm{\eta},\bm{\eta'})=\left<\psi^\dagger(\bm{\eta})\psi(\bm{\eta'})
\right>$.  In Eq.~(\ref{eq:g22g11g11}), the positive (negative) sign imply the
bunching (antibunching) effect for bosons (fermions).  Further , after
introducing a quantity named intensity fluctuation, $\Delta
I(\bm{\xi})=I(\bm{\xi})-\left<I(\bm{\xi})\right>$, from
Eqs.~(\ref{eq:g22})-(\ref{eq:g22g11g11}) the correlation between the intensity
fluctuations at the two detectors becomes\cite{HSS-PRL2004}
\begin{eqnarray}
&&\left<\Delta I_r(\bm{\xi}_r)\Delta I_t(\bm{\xi}_t)\right>=\nonumber\\
&&\quad\pm v_iv_f\left|\int d\bm{\eta}d\bm{\eta'}
G^{(1,1)}(\bm{\eta},\bm{\eta}')h^*_r(\bm{\xi}_r,\bm{\eta})h_t(\bm{\xi}_t,
\bm{\eta}')\right|^2.
\label{eq:deltaII}
\end{eqnarray}
Here, the initial and final velocities $v_i$ and $v_f$ are included for massive
particles.

We consider an incident neutron field with a fully spatially
incoherent wavefront,
\begin{equation}
G^{(1,1)}(\bm{\eta},\bm{\eta}')=\left<\varphi^\dagger(\bm{\eta})\varphi(
\bm{\eta}')\right>=I_0\delta(\bm{\eta}-\bm{\eta}'), 
\label{eq:g11delta}
\end{equation}
where $I_0$ is the beam intensity per unit area.  In fact, it is this pattern
that encodes the wavefront and makes coincidence imaging possible.  With
Eq.~(\ref{eq:g11delta}), the calculation of Eq.~(\ref{eq:deltaII}) is
essentially reduced to the derivation of $h_r(\bm{\xi}_r,\bm{\eta})$ and
$h_t(\bm{\xi}_t,\bm{\eta})$.

We consider quasimonochromatic and noninteracting thermal neutrons with spin
polarized in the $\hat{z}$ direction.  The coordinate system is defined in
Fig.~\ref{fig1}b.  The incident neutron wave function after the 
beamsplitter\cite{*[{See Supplemental Material at }][{ for details.}] supplemental1}
\begin{equation}
\psi_{\text{in}}(\mathbf{r})=\frac{1}{i\lambda}\int d\bm{\eta}\,\varphi(
\bm{\eta})\frac{\exp\left(ik\left|\mathbf{r}-\mathbf{r}_0\right|\right)}
{\left|\mathbf{r}-\mathbf{r}_0\right|}\begin{bmatrix} 1\\ 0 \end{bmatrix}
\label{eq:psiin}
\end{equation}
is an exact solution to the Schr\"{o}dinger equation with $\varphi(x_0,z_0)$ as
the field source, where $\bm{\eta}=(x_0,z_0)$, and $\lambda$ and $k$ are the
wavelength and wavevector, respectively.  Immediately, the reference wave is
obtained as $\psi_{\text{in}}(\mathbf{r}_r)$.

However, calculating the neutron wave function at the target detector is
challenging.  Conventional neutron scattering theory is established on plane
wave incidence and invalid for the incidence field in Eq.~(\ref{eq:psiin}).
The interaction between neutron and matter is best described as a potential
scattering problem with $V(\mathbf{r'})=\sum_j V_j^n(\mathbf{r'})+\sum_i
V_i^m(\mathbf{r'})$ as the potential\cite{Squires1978}.  The first summation
is over all nuclear sites with the Fermi pseudopotential for the $j$-th nucleus
\begin{equation}
V_j^n(\mathbf{r'})=\frac{2\pi
\hbar^2}{m_n}\left[A_j+B_j\bm{\sigma}\cdot\mathbf{I}_j\right]
\delta(\mathbf{r'}-\mathbf{r}_j),\label{eq:Vnuclear}
\end{equation}
while the second summation is over all unpaired electron sites with the magnetic
potential for the $i$-th electron
\begin{eqnarray}
V_i^m(\mathbf{r'})&=&-\frac{\hbar^2}{2m_n}\gamma
r_e\bm{\sigma}\cdot\left[\mathbf{W}_{Si}(\mathbf{r'})+
\mathbf{W}_{Li}(\mathbf{r'})\right],\label{eq:Vmagnetic}\\
\mathbf{W}_{Si}(\mathbf{r'})&=&\nabla\times\left(\frac{\mathbf{s}_i\times
\mathbf{\hat{R}}}{R^2}\right),\ 
\mathbf{W}_{Li}(\mathbf{r'})=\frac{1}{\hbar}\frac{\mathbf{p}_i\times
\mathbf{\hat{R}}}{R^2}.\label{eq:Wi}
\end{eqnarray}
In the above equations, $m_n$ is the neutron mass, $r_e$ the classical radius of
electron, $\gamma=1.913$; $A_j=\left[\left(I_j+1\right)b_j^++I_jb_j^-\right]/
\left(2I_j+1\right)$, $B_j=\left[b_j^+-b_j^-\right]/\left(2I_j+1\right)$, $I_j$
is the nuclear spin of the $j$-th nucleus.  The neutron plus nucleus can form
total spins $I_j+\frac{1}{2}$ and $I_j-\frac{1}{2}$, and $b_j^+$ and $b_j^-$ are
the free scattering lengths of the two corresponding eigenstates, respectively.
Because coherent nuclear scattering does not change neutron spin, we drop the
$B_j$ term in this paper to keep our discussion focused.   The Pauli matrix
$\bm{\sigma}$ in Eq.~(\ref{eq:Vmagnetic}) is the operator for the neutron's
spin.  The magnetic field contains two contributions.  $\mathbf{W}_{Si}$ arises
from the electron's magnetic moment, whereas $\mathbf{W}_{Li}$ originates from
the electron's orbital movement.  $\mathbf{s}_i$ and $\mathbf{p}_i$ are
electron's spin and momentum operators, respectively; $\mathbf{\hat{R}}$ is the
unit vector of $\mathbf{R}$ ($\equiv \mathbf{r'}-\mathbf{r}_i$) with
$\mathbf{r}_i$ the electron position.

For simplicity, we restrict our discussion in this paper to elastic scattering
only.  The following derivation can be easily generalized to include inelastic
scattering.  Elastic vs inelastic scattering is routinely resolved by means of
energy-analyzing crystal at the detector side, where neutrons carrying the
incident energy correspond to elastic scattering.  We start from the
Lippmann-Schwinger integral equation\cite{Cohen-Tannoudji2005}, i.e.
\begin{equation}
\psi(\mathbf{r})=\psi_0(\mathbf{r})+\frac{2m_n}{\hbar^2}\int d^3r'
G(\mathbf{r}-\mathbf{r'})V(\mathbf{r'})\psi(\mathbf{r'}),
\label{eq:Lippmann}
\end{equation}
where $\psi_0(\mathbf{r})$ is a solution of the homogeneous equation
$(\nabla^2+k^2)\psi_0(\mathbf{r})=0$ and $G(\mathbf{r}-\mathbf{r'})$ is the
outgoing Green's function given by
\begin{equation}
G(\mathbf{r}-\mathbf{r'})=-\frac{\exp(ik\left|\mathbf{r}-\mathbf{r'}\right|)}
{4\pi\left|\mathbf{r}-\mathbf{r'}\right|}.
\label{eq:Green}
\end{equation}
When the potential $V$ is considered a small perturbation, the $\psi$ in the
integrand can be approximated by the incident wave function $\psi_{\text{in}}$.
Due to $(\nabla^2+k^2)\psi_{\text{in}}(\mathbf{r})\propto\delta(y-y_0)$, 
when the sample is far away from the beamsplitter, coordinate $y_0$ is considered outside the
sample space.  Therefore, $\psi_{\text{in}}(\mathbf{r})$ sufficiently satisfies
the homogeneous equation and $\psi_0(\mathbf{r})$ can be replaced by
$\psi_{\text{in}}(\mathbf{r})$ as well.  The integral term on the rhs of
Eq.~(\ref{eq:Lippmann}) is identified as the scattering wave function.

In Fig.~\ref{fig1}a polarized neutrons are collected.  Under the diffuse
illumination (Eq.~(\ref{eq:g11delta})), neutrons can fly in random directions.  Thus for spin-up detection,
an extra beam-stop screen is required to remove the incidence wave, and the
sample is mounted in the small opening window (Fig.~\ref{fig1}a).  Substituting
Eqs.~(\ref{eq:Vnuclear})-(\ref{eq:Wi}) and (\ref{eq:Green}) into
Eq.~(\ref{eq:Lippmann}), under the only assumption that the sample size is much
smaller than the source-sample distance $d_1$ and the sample-detector distance
$d_2$, the scattering wave function can be expressed as\cite{supplemental1}
\begin{eqnarray}
\psi^{\text{sc}}_{\nu'\nu}(\mathbf{r})&&=\frac{i}{\lambda}\int d\bm{\eta}
\varphi(\bm{\eta})\int d^3r'\frac{\exp[ik(r_1+r_2)]}{r_1r_2}\nonumber\\
&&\ \left\{\beta\bm{\sigma}\cdot\left[\bm{\hat\kappa}_c\times\left(\mathbf{M}
(\mathbf{r'})\times\bm{\hat\kappa}_c\right)\right]+A(\mathbf{r'})\right\}
\begin{bmatrix}1\\ 0\end{bmatrix},
\label{eq:psisc}
\end{eqnarray}
where $\left|\nu\right>$ and $\left|\nu'\right>$ are the initial and final
states of the sample, $A(\mathbf{r'})=\left<\nu'\left|\sum_j 
A_j\delta(\mathbf{r'}-\mathbf{r}_j)\right|\nu\right>$,$\mathbf{M}(\mathbf{r'})=
\left<\nu'\left|\bm{\mathcal{M}}(\mathbf{r'})\right|\nu\right>$, 
$\bm{\mathcal{M}}(\mathbf{r'})$ is the operator for the local magnetization,
$\beta=\gamma r_e/(2\mu_B)$, $\mathbf{r}_1(\mathbf{r'},\mathbf{r}_0)
=\mathbf{r'}-\mathbf{r}_0$, $\mathbf{r}_2(\mathbf{r},\mathbf{r'})
=\mathbf{r}-\mathbf{r'}$, $\bm{\kappa}_c=k\left[\mathbf{\hat
r}_1(\mathbf{r}_c,\mathbf{r}_0)-\mathbf{\hat r}_2(\mathbf{r},\mathbf{r}_c)
\right]$, and $\mathbf{r}_c$ is the center of the sample and $\mathbf{r}_0$ a
point on the source.  $\bm{\kappa}_c$ varies when the source point $\mathbf{r}_0$
changes.

Now the $h_r$ and $h_t$ functions are readily extractable from
Eqs.~(\ref{eq:psiin}) and (\ref{eq:psisc}).  We emphasize that Eq.~(\ref{eq:psisc})
is valid for all sizes of source and scattering angle $\theta$.  In the
following discussion, we would focus on the case of small angle scattering with
a beam size much smaller than $d_1$ and $d_2$ so that the paraxial approximation
is applicable.  The scattering vector $\bm{\kappa}_c$ can now be substituted with
the conventional $\bm{\kappa}_0$, whereas varations of $\bm{\kappa}_c$ from
$\bm{\kappa}_0$ only make negligible corrections.  Consequently, $\bm{\kappa}_0$
can be taken out of the integrand in Eq.~(\ref{eq:psisc}).  We would like to
define the probed sample functions as 
\begin{equation}
\begin{bmatrix}S^\uparrow(\mathbf{r'})\\
S^\downarrow(\mathbf{r'})\end{bmatrix}=\left\{\beta\bm{\sigma}\cdot\left[
\bm{\hat\kappa}_0\times\left(\mathbf{M}(\mathbf{r'})\times\bm{\hat\kappa}_0\right)\right]
+A(\mathbf{r'})\right\}\begin{bmatrix}1\\0 \end{bmatrix}.
\label{eq:sample}
\end{equation}
Under paraxial condition, we have
\begin{equation}
h_r(\bm{\xi}_r,\bm{\eta})=\frac{e^{ikd_r}}{i\lambda
d_r}\exp\left[\frac{i\pi}{\lambda
d_r}\left(\bm{\xi}_r-\bm{\eta}\right)^2\right].
\label{eq:hrparaxial}
\end{equation}
The impulse response function for the sample arm is
\begin{equation}
h_t^p(\bm{\xi}_t,\bm{\eta})=\frac{i}{\lambda}\int d^3r'\frac{\exp\left[
ik(r_1+r_2)\right]}{r_1r_2}S^p(\mathbf{r'})
\label{eq:htparaxial}
\end{equation}
with spin index $p=\uparrow,\downarrow$.  Substituting Eqs.~(\ref{eq:g11delta}),
(\ref{eq:hrparaxial})-(\ref{eq:htparaxial}) into Eq.~(\ref{eq:deltaII}) and
selecting $d_r=d_1+d_2$, a direct integration generates a very simple form,
\begin{eqnarray}
&&\left<\Delta I^\uparrow_r(\bm{\xi}_r)\Delta I_t^p(\bm{\xi}_t)\right>\nonumber\\
&=&-\sum_{\nu,\nu'}p_\nu\,\chi\left|\int d\bm{\zeta}\exp\left[\frac{i2\pi\left(
\bm{\xi}_r-\bm{\xi}_t\right)\cdot\bm{\zeta}}{\lambda d_2}\right]\int dy'S^p(
\mathbf{r'})\right|^2\nonumber\\
&=&-\sum_{\nu,\nu'}p_\nu\,\chi\left|\mathcal{F}\left[\mathcal{P}S^p\right]\left(
\frac{2\pi\left(\bm{\xi}_r-\bm{\xi}_t\right)}{\lambda d_2}\right)\right|^2,
\label{eq:FourierS}
\end{eqnarray}
where $\mathcal{P}$ is the longitudinal projection operation along the $y$-axis,
i.e., $\mathcal{P}S^p(\bm{\zeta})\equiv \int dy'S^p(\mathbf{r'})$ and
$\bm{\zeta}=(x',z')$; $\mathcal{F}$ is the 2d Fourier transform on the
transverse plane. The consecutive operations $\mathcal{F}$ and $\mathcal{P}$
convert a function in 3d real space to a function in 2d $k$-space.  The
parameter $\chi=4\pi^2\hbar^2I_0^2/(\lambda^4d_2^4m_n^2)$ and we have
substituted the neutron velocity $v=\hbar k/m_n$ into Eq.~(\ref{eq:deltaII}).
In Eq.~(\ref{eq:FourierS}) we have written down the average over the sample's
initial state $\left|\nu\right>$ and the summation over the final state
$\left|\nu'\right>$ explicitly, where $p_\nu$ is the initial state's
distribution.  The isotope averages are excluded.  The isotope structural
information should be one of the imaging goals.

The experimental signal on the lhs of Eq.~(\ref{eq:FourierS}) only gives the 
amplitude 
of the Fourier transform of $\mathcal{P}S^p(\bm{\zeta})$.  An inverse Fourier
transform would still require the phase information.  Phase-retrieval
technique has been an intensively-studied area in applied mathematics, and a
number of sophiscated algorithms have been developed\cite{Shechtman-IEEE2015}.
This allows retrieval of the Fourier phase from the Fourier amplitude alone.
The first ground breaking application of phase-retrieval was achieved in X-ray
imaging\cite{MiaoJW-Nature1999,MiaoJW-PRL2002} and later successively employed
in many areas including ghost imaging\cite{HSS-PRL2016}.  Phase-retrieval is
alread a standard part of diffraction imaging.  Here, with this powerful tool,
the image of $\mathcal{P}S^p(\bm{\zeta})$ in real space can be reconstructed.
From Eq.~(\ref{eq:sample}) the probed sample functions $S^\uparrow(\mathbf{r'})$ and
$S^\downarrow(\mathbf{r'})$ are linear combinations of $M_x(\mathbf{r'})$,
$M_y(\mathbf{r'})$, $M_z(\mathbf{r'})$ and $A(\mathbf{r'})$.  Obtaining
individual functions of $M$'s and $A$ from $S$'s would require multiple linearly independent equations.
Fortunately, by placing the sample detector $D_t$ at different locations,
satisfaction of this condition is guaranteed\cite{supplemental1}.  For example,
we consider three detector locations $(x_t,z_t)$: position 1 $(\xi,0)$, position
2 $(-\xi,0)$, and position 3 $(0,\xi)$ with $\xi=d_2sin\theta$.  The independent
equations would be
\begin{equation}
\begin{bmatrix}S_1^\uparrow\\ S_1^\downarrow\\
S_2^\downarrow\\ S_3^\uparrow\\
S_3^\downarrow\end{bmatrix}=
\begin{bmatrix}
0 & 0 & 1 & 1\\
ie^{-i\frac{\theta}{2}}\sin\frac{\theta}{2} &
ie^{-i\frac{\theta}{2}}\cos\frac{\theta}{2} & 0 & 0\\
-ie^{i\frac{\theta}{2}}\sin\frac{\theta}{2} &
ie^{i\frac{\theta}{2}}\cos\frac{\theta}{2} & 0 & 0 \\
0 & \frac{1}{2}\sin\theta & \sin^2\frac{\theta}{2} & 1 \\
1 & i\cos^2\frac{\theta}{2} & \frac{i}{2}\sin\theta & 0
\end{bmatrix}
\begin{bmatrix} \beta M_x\\ \beta M_y\\ \beta M_z\\ A \end{bmatrix}.
\label{eq:SMA}
\end{equation}
We pursue $M$'s and $A$ as the overdetermined solution to Eq.~(\ref{eq:SMA}) to
avoid possible ill behaved results due to small angle $\theta$.  We emphasize
that Eq.~(\ref{eq:SMA}) is only a representative example and other optimized
combinations are possible.  Eqs.~(\ref{eq:FourierS}) and (\ref{eq:SMA})
together lead to individual functions of $\mathcal{P}M_x$,
$\mathcal{P}M_y$, $\mathcal{P}M_z$ and $\mathcal{P}A$.  These projection
functions are 2d transverse images.  By rotating the sample around the $z$-axis
or the $x$-axis, multiple projections can be obtained and the
x-ray CT algorithms can be employed.  This enables a
3d tomographic image reconstruction of the sample.

Based on Eq.~(\ref{eq:FourierS}), the intensity coincidence imaging system
(Fig.~\ref{fig1}a) actually achieves the function of Fourier-transform imaging
without any optical instruments (such as lens) in the setup.  There are two
critical elements in this imaging scheme at the source part, the antibunching
inherited from the Pauli exclusive principle and the totally incoherent
wavefront (in ensemble average sense).  They together determine the
characteristic statistics of the incident field.  Polarized neutrons
should be used, not only because it would simplify the analysis, but also the
antibuching contrast of unpolarized neutrons is only half of that of polarized
neutrons\cite{Iannuzzi-PRL2006}.  The signal at the sample detector $D_t$ is the
summation of the contributions from different portions of the wavefront.  The
second element prevents interference between these contributions
(Eq.~(\ref{eq:g11delta})) and ensures the transverse coordinates be encoded by
the wavefront.  The resolving power of this imaging technique originates from
this special wavefront.

Unlike in optical GI where semi-classical treatment is possible, the fermionic
antibunching is purely quantum mechanical.  In this paper we start from the
first principle of quantum mechanics and establish our analysis on the firm
basis of Schr\"{o}dinger equation and the equivalent Lippmann-Schwinger
equation.  All results naturally flow out from the calculations.  What physical
quantities are imaged are distinctly clear.  Previously, second-order fermionic
GI was only attemped in Ref.~\cite{WangKG-PRA2009}, with a follow-up study on
higher-order fermionic correlations\cite{LiuHC-PRA2016}.  In these
works, fermions and bosons were handled in the same fashion as scalars, and the
only difference between each other is the spatial symmetry vs anti-symmetry in
the source field statistics.  However, this treatment is not only inadequate,
but also problematic, as bunching and antibunching are spatial effects.  Under
certain circumstances, spin-spatial combination induces two-fermion bunching.
For example, unpolarized fermions (spin 1/2) still exhibit an overall
antibunching effect\cite{Iannuzzi-PRL2006}, with 25\% bunching component and
75\% antibunching component.  The two correlated fermions can form either a
state with total spin $S=0$ (bunching), or states with total spin $S=1$ and
$S_z=+1,0,-1$ (antibunching).  A scalar theory cannot be a truly correct one,
and fails at least in the two-fermion bunching scenario.  GI based on
unpolarized thermal neutron antibunching can provide valuable information in
various situations such as symmetry breaking in the target.  Spinor
representation for fermions (Eq.~(\ref{eq:psiin})) is not only an additional
dimension and degree of freedom, but also a necessity to describe the full
spin-spatial behavior, let alone required by quantum mechanics.  Therefore, our
work provides the first, rigorous, truly fermionic GI scheme.
 
In summary, we present a lensless, Fourier-transform imaging scheme
based on intensity correlations of thermal neutrons.  As coherent neutron
source is unavailable, high order coherence embedded in neutron intensity is
explored.  Such coherence is inherited from the Pauli exclusion of fermions and
intrinsic in the quantum statistical properties of neutron field.  The
interaction between the thermal neutron field and the sample's nuclear sites
and internal magnetic structure is studied and a tomographic
reconstruction technique proposed.  There is no requirement for periodicity. So
crystallined or noncrystallined structures can be investigated.  Essentially an
intensity interference technique without necessity for phase coherence, high
brightness fluxes from nuclear reactor or spallation sources can be employed.
In addition, by avoiding aberration problems of lens systems, Fourier-transform
imaging can typically achieve resolution at the wavelength level.  Our
development opens new avenues for high-resolution, high throughput quantum
microscopy of matter.  Especially, the capability of resolving micro magnetic
structures within the interior of samples is a unique feature of our GI.  This
will solve problems in many research frontiers.  For example, metalloproteins
comprise approximately half of all proteins, and the metal sites often determine
the protein functions\cite{Zatta2004,Messerschmidt2004}.  Any sites containing
unpaired electrons become good magnetic targets.  Resolving these structures
will provide invaluable information for studying the dynamics of protein
functioning.  Technology originated from our work would attract tremendous
interests and lead to enormous applications in condensed matter, material
science and biostructural analysis.

\begin{acknowledgments}
This work was supported by the National Natural Science Foundation of
China under Grant Project No. 11627811, the National Key Research and
Development Program of China under Grant No. 2017YFB0503303.
\end{acknowledgments}

\providecommand{\noopsort}[1]{}\providecommand{\singleletter}[1]{#1}%

\end{document}